\documentclass[prb,twocolumn,aps,showpacs]{revtex4}
\usepackage{hyperref}
\usepackage{bm}
\def\ltsim{\vbox {\hbox{\lower .8\baselineskip \hbox{$<$}} \break
                 \hbox{\lower 0.2\baselineskip \hbox{$\sim$}} } }

\begin{document}

\title{Hidden Caldeira-Leggett dissipation in a Bose-Fermi Kondo model}
\author{Mei-Rong Li$^1$, Karyn Le Hur$^1$, and Walter Hofstetter$^2$}
\affiliation{$^1$D\'epartement de Physique and RQMP, Universit\'e de Sherbrooke, 
Sherbrooke, Qu\'ebec, Canada J1K 2R1\\
$^2$ Institut f\"ur Theoretische Physik A, RWTH Aachen,  Templergraben 55,
52056 Aachen, Germany}

\begin{abstract}
We show that the Bose-Fermi Kondo model (BFKM), which may find applicability both 
to certain dissipative mesoscopic qubit devices and to heavy fermion systems described 
by the Kondo lattice model, can be mapped exactly onto the Caldeira-Leggett model. This 
mapping requires an ohmic bosonic bath and an Ising-type coupling between the latter 
and the impurity spin. This allows us to conclude unambiguously that there is an 
emergent Kosterlitz-Thouless quantum phase transition in the BFKM with an ohmic 
bosonic bath. By applying a bosonic numerical renormalization group approach, 
we thoroughly probe physical quantities close to the quantum phase transition.   
\end{abstract}

\maketitle

The Bose-Fermi Kondo model (BFKM) (or equivalently the spin-boson-fermion model),
originally introduced by Si and coworkers \cite{QSi} and by Sengupta \cite{Sengupta} 
to describe peculiar quantum critical behaviors in heavy-fermion Kondo lattice 
systems\cite{Lohneysen}, involves a single impurity spin being coupled both to a 
{\it bosonic} bath and to a {\it fermionic} bath. Resulting from the nontrivial 
competition between these distinct baths, a rich phase diagram is known to emerge 
from this model (See, {\em e.g.}, Refs.~\onlinecite{Si} and \onlinecite{ZD}). More 
generally, a great interest is currently devoted to the understanding of the Kondo 
entanglement breakdown mechanism due to the presence of extra (here, bosonic) quantum 
fluctuations resulting in striking quantum phase transitions\cite{Ingersent}. In this 
Letter, we revisit the case where the impurity spin $S=1/2$ is coupled to an 
{\em ohmic} bosonic bath with a continuum spectrum -- ohmic means that 
the bosonic correlation function in time $t$ decays as $1/t^2$ -- through an Ising 
coupling. The anisotropic Hamiltonian under consideration thus takes the form:
\begin{eqnarray}
& & H = hS_z+ H_{sf}+H_{sb},   \label{BFK}\\
& & H_{sf} = {J_{\perp}\over 2}\left(\Psi^{\dagger}_{\downarrow}(0)
\Psi_{\uparrow}(0)S_+ +{\rm h.c.}\right)  \nonumber \\
& &\;\;\;\;\;\;\;\;\;\;\; +v_f \sum_{\sigma=\uparrow,\downarrow} 
\int^{\infty}_{-\infty} 
dx\ \Psi^{\dagger}_{\sigma}(x) i\partial_x \Psi_{\sigma}(x), \label{Hsf} \\
& &H_{sb} =-{v_b\partial_x\Phi(0)\over \sqrt{2 K_b}} \, S_z  + 
{v_b\over 4\pi} \int^{\infty}_{-\infty} dx [\partial_x \Phi(x)]^2,
\label{Hsb}
\end{eqnarray}
where $h$ is a magnetic field, $\Psi_\sigma(x)$ and $\Phi$ represent 
the fermionic and bosonic fields; $v_b$ is the velocity of the bosons and $K_b^{-1}\neq 0$
the typical coupling between the bosons and the impurity spin; Recall that 
$K_b^{-1}=0$ means no coupling between the impurity spin and the bosonic 
environment. In Eq.~(\ref{Hsf}), $v_f$ is the Fermi velocity. Without loss of generality, the 
one-dimensional character of the Hamiltonian $H_{sf}$ can be viewed as a result 
of the point-like character of the impurity and the rotational symmetry. 
Besides, we omit the Ising part of the Kondo coupling $J_z$ 
due to its minor effect (see the last page).

Our interest in the model (\ref{BFK}-\ref{Hsb}) is also motivated by the 
fact that this model may be realized in mesoscopic dissipative setups 
involving qubits, as pointed out by one of us recently\cite{karyn}. 
More precisely, 
the impurity spin can embody the two allowed charge states of a big metallic grain 
close to a given degeneracy point and $h$ being proportional to the gate voltage 
measures deviations from this degeneracy point\cite{Matveev}. The conduction electrons stand 
for the electrons both in the metallic grain $(\Psi_{\downarrow})$ and in a nearby 
reservoir electrode $(\Psi_{\uparrow})$, and the $J_{\perp}$ (Kondo) term denotes the 
tunneling process of an electron from lead to grain that flips (through the raising 
operator $S_+$) the charge state of the grain, and vice-versa. Note that here the 
spin index $\sigma=(\uparrow,\downarrow)$ is completely artificial and refers to the 
position of an electron in the structure (lead or grain). The original spin of the 
electrons are assumed to be polarized (``spinless electrons'') due to the application 
of a strong magnetic field. The bosons represent the electromagnetic noise in the 
gate voltage stemming from the finite resistance $R$ in the gate lead, and the 
spin-boson coupling reflects the effect of the voltage noise on the charge 
fluctuations of the grain. Here, $K_b=R_K/2R$ with $R_K=2\pi\hbar/e^2$ the quantum 
of resistance\cite{karyn}. Two of us have extended this model to one-dimensional reservoir 
leads being embodied by a Luttinger liquid behavior\cite{LL2} and also to two strongly 
capacitively-coupled large quantum dots (which can be reduced to a single 
charge qubit)\cite{LL1}.

As a first step, the model of Eqs.~(\ref{BFK}-\ref{Hsb}) has been extensively studied 
\cite{Si,ZD,karyn} by using a perturbative renormalization group (RG) approach. A 
quantum phase transition is discovered at the critical value of $K_b$ 
\cite{karyn,note2}:
\begin{eqnarray}
(K_b)_c^{-1} = J_\perp/\pi v_f = 2\Delta/\omega_c,         \label{pRG}
\end{eqnarray}
where $\omega_c$ is the high energy cutoff in the theory and $\Delta=J_\perp \omega_c/2\pi v_f$.
The RG flow equations suggest that the phase transition is of the Kosterlitz-Thouless 
(KT) type. On the other hand, a more rigorous approach is required in order to unambiguously prove
the KT transition as well as to scrutinize the evolution of physical quantities in the vicinity of 
the quantum phase transition. In this Letter, we explore an exact mapping of the BFKM in 
Eqs.~(\ref{BFK}-\ref{Hsb}): We properly demonstrate that this model can be 
mapped onto the Caldeira-Leggett (CL) model \cite{CL} with the effective dissipation 
strength
\begin{eqnarray}
\label{alpha}
\alpha = 1+(4 K_b)^{-1},  \label{alphaKb}
\end{eqnarray}
and therefore the two models belong to the same class of universality. It immediately
follows that there is a KT quantum phase transition in the BFKM,
separating a Kondo phase and an unscreened spin phase.  Moreover, it irrefutably demonstrates the 
important Eq. (4). Even though the Bethe-Ansatz method\cite{BA} can be exploited in the Kondo region, 
it breaks down in the vicinity of the quantum phase transition\cite{Buttiker} (which must be 
identified as the antiferromagnetic-ferromagnetic transition in the anisotropic Kondo model). We 
resort to a bosonic numerical RG (NRG) technique applying to the CL model\cite{Bulla}. Of interest 
to us  here is the spin magnetization at temperature $T=0$, $\langle S_z \rangle$,
as well as the (local) spin susceptibility $\chi_{\rm loc}(T)$ versus $T$.
In the mesoscopic realizations, $\langle S_z \rangle$ = $\langle Q \rangle-1/2$,
where $Q$ is the charge operator on the (large) dot, can be measured with very high 
precision\cite{karyn,LL2}. 

{\em CL mapping.}---We first bosonize the fermions in Eq.~(\ref{Hsf}) as: 
$\Psi^{\dagger}_{\downarrow}(x)\Psi_{\uparrow}(x) = [\omega_c/(2\pi v_f)]
\exp(i\sqrt{2}\varphi(x))$ and $\Psi^{\dagger}_{\uparrow}(x)\Psi_{\uparrow}(x)
-\Psi^{\dagger}_{\downarrow}(x)\Psi_{\downarrow}(x) =
\partial_x \varphi(x)/(\sqrt{2}\pi)$; $\Delta$ and $\omega_c$ are defined in Eq. (4). 
By dropping the trivial contribution in the charge sector,
the Hamiltonian $H_{sf}$ becomes
\begin{eqnarray}
H_{sf} = {\Delta\over 2} \left[e^{i\sqrt{2}\varphi(0)} S_+ + h.c.\right] 
+ {v_f\over 4\pi} \int^{\infty}_{-\infty} dx \big[\partial_x \varphi(x)\big]^2, 
\label{Hsf1}
\end{eqnarray}
and the Hamiltonian (\ref{BFK}) becomes that of an impurity spin coupled to {\em two} 
bosonic baths. Now we show that in the case of an ohmic bosonic bath in Eq. (3) one 
can envision to combine the two bosonic baths and eventually map the 
model onto a well-known CL model. The local action for the 
boson fields $\varphi(x)$ reads  $S^{loc}_{\varphi} = {T\over 2\pi} 
\sum_{\omega_n} |\omega_n| \varphi_0(\omega_n)\varphi_0(-\omega_n)$,
where $\varphi_0=\varphi(x=0)$ and $\omega_n$ is the bosonic Matsubara frequency. 
Such a form implies that $\varphi(x)$ is of the ohmic type.
If the bosonic bath $\Phi(x)$ in Eq.~(\ref{Hsb}) is also ohmic and thus embodied by 
the local action $S^{loc}_{\Phi} =  {T\over 2\pi} \sum_{\omega_n}  
|\omega_n| {\Phi}_0(\omega_n){\Phi}_0(-\omega_n)$ with $\Phi_0=\Phi(x=0)$, 
we can then employ linear combinations of $\varphi_0$ and ${\Phi}_0$ to 
simplify the problem.
(Note that starting from a non-ohmic  bosonic environment \cite{Grimpel} instead, 
one would replace  $|\omega_n|$ in $S^{loc}_{\Phi}$ by $|\omega_n|^{s}$
with $s\neq 1$; the two bosonic baths thus would have different dynamics which 
would hinder the straightforward combination.)  By performing the unitary transformation
${\cal U}_1=\exp\{-i\Phi_0S_z/\sqrt{2 K_b}\}$ to Eqs.~(\ref{BFK}), (\ref{Hsb})
and (\ref{Hsf1}), we can absorb the spin-boson coupling term in  Eq.~(\ref{Hsb})
into the transverse Kondo term $\Delta$ in Eq.~(\ref{Hsf1}).
It is then convenient to introduce the symmetric and antisymmetric bosonic 
combinations $\varphi_s = (2\alpha)^{-1/2} [\sqrt{2} \varphi_0 + (2K_b)^{-1/2} \, 
\Phi_0]$ and $\varphi_a = (2\alpha)^{-1/2}[(2K_b)^{-1/2}\, \varphi_0 -\sqrt{2}\Phi_0]$, 
where the crucial parameter $\alpha$ has been defined in Eq.~(\ref{alphaKb}). 
The impurity spin gets only coupled to the $\varphi_s$ mode and it is enough 
to write down the $\varphi_s$-action
\begin{eqnarray}
 S_{\varphi_s} &=& {T\over 2\pi} 
\sum_{\omega_n} |\omega_n| \varphi_s(\omega_n) \varphi_s(-\omega_n) \nonumber \\ 
&-&{\Delta\over 2}\, \big[e^{ i \sqrt{2\alpha} {\varphi_s}(0)} S_+ +h.c.\big].
\end{eqnarray}
Conceptually, we can visualize $S_{\varphi_s}$ as the action linked to an 
Hamiltonian $H_{\varphi_s}$ with the effective velocity $v_s=\omega_c a$ ($a$ is the 
short-distance cutoff which will be used below) and modeling a single bosonic bath with 
the dissipative parameter $\alpha$ coupled to the impurity spin. The link with the CL 
model of a two-level system with ohmic dissipation 
 \cite{CL} becomes clear when 
applying the unitary transformation ${\cal U}_2=\exp\{i \sqrt{2\alpha} \varphi_s(0)S_z\}$ 
to turn the $\Delta$ term into a transverse magnetic field $\Delta S_x$, leading to 
\begin{eqnarray}
\label{CL}
\widetilde{H}=H_S -\sqrt{2\alpha} v_s 
\partial_x \varphi_s(0) S_z 
+\frac{v_s}{4\pi} \int^\infty_{-\infty} dx [\partial_x \varphi_s(x)]^2,
\end{eqnarray}
where $H_S=h S_z +{\Delta} S_x$ and $\alpha$ is given in Eq. (\ref{alphaKb}).

This mapping is crucial because it irrefutably reveals an originally hidden 
correspondence between the BFKM of Eqs.~(1-3) and the CL model in which the breakdown 
of the Kondo physics (quantum coherence) due to dissipation 
is clearly established. For example, a KT type transition was predicted in 
Ref.~\onlinecite{CL} in the context of ohmic dissipation and this has been solidly 
confirmed through the NRG, either by refermionizing the bosonic degrees of freedom 
and then mapping the CL model onto the anisotropic Kondo model\cite{costi}, or  
directly by resorting to the bosonic NRG\cite{Bulla}. For $\Delta\rightarrow 0$ the 
KT transition is known to occur at \cite{CL} $\alpha=\alpha_c=1$ implying in the BFKM 
a critical $(K_b)^{-1}_c\rightarrow 0$ from Eq.~(\ref{alphaKb}): since 
$\Delta\rightarrow 0$, the Kondo energy scale $T_K$ characterizing the emergence of 
a bound state between the local moment and the conduction electrons vanishes and 
thus the local moment remains unscreened whatever the coupling between the 
dissipative mode $\partial_x\Phi(0)$ and the local spin. 

To understand more deeply the connection between the CL model and the Kondo phenomenon, 
we can proceed from Eq. (8) along the lines of Ref. \onlinecite{Furusaki} and precisely 
recover the anisotropic Kondo model. The effective Kondo parameters here are given by 
$J_{\perp}=\pi v_s \Delta/\omega_c$ and $J_z=4\pi v_s(1-\sqrt{\alpha})$. Thus, one can 
clearly distinguish two phases; a Kondo (delocalized) realm when $J_z>-|J_{\perp}|$ and 
a ferromagnetic Kondo phase which embodies an unscreened (localized) moment when 
$J_z<-|J_{\perp}|$. This enables us to rigorously predict a KT phase transition in the 
BFKM when $J_z=-|J_{\perp}|$ which implies  a critical $\alpha_c=1+(K_b)_c^{-1}/4=
1+\Delta/(2\omega_c)$ when assuming $0<\Delta\ll \omega_c$; this reinforces the 
intuition gained from the perturbative RG analysis of the BFKM\cite{karyn}.  Since 
Bethe-Ansatz calculations\cite{BA} can only be applied to this model when
$\alpha<\alpha_c$ \cite{Buttiker}, we seek to apply the (bosonic) NRG on Eq.~(8) to 
probe $\langle S_z \rangle$ versus $h$ and $\chi_{loc}(T)$ very close to 
$\alpha_c$. Those quantities have not been studied in Ref. \onlinecite{Bulla}.

\begin{figure}[ht]
\begin{picture}(180,98)
\leavevmode\centering\includegraphics{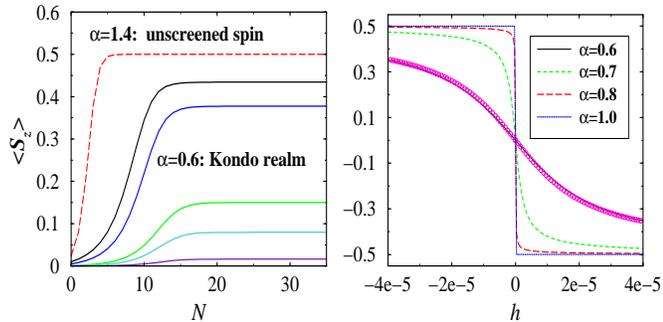}
\end{picture}
\caption{(color online) Left: Flow of $\langle S_z \rangle$ with the iteration $N$. 
The dashed line is for $h=-0.0001$ (unscreened realm). All the solid lines correspond 
to $\alpha=0.6$ (Kondo regime), and from top to bottom, to $h=-10^{-4}, -5\times 
10^{-5}, -10^{-5}, -5\times 10^{-6}, -10^{-6}$. Right: $\langle S_z \rangle$ as a 
function of $h$. The open circles sitting on top of the $\alpha=0.6$ curve are the 
Bethe-Ansatz results\cite{Buttiker,BA}. Parameters are $\Delta=0.01$ and $\omega_c=1$. 
The NRG parameters are $\Lambda=2, N_s=100, N_b=8$, and $N_{b0}=500$. } 
\label{Sz}
\end{figure}

{\em NRG endeavors close to the transition}.---The strength of the NRG lies in its 
nonperturbative nature and the ability to resolve arbitrarily small energies
\cite{Wilson,Krishna,Hofstetter}. This allows to provide important information 
in the phase transition region. For convenience, the Hamiltonian (\ref{CL}) can be 
rewritten as\cite{note} 
\begin{eqnarray}
\widetilde{H}=H_S +\omega_c \int^1_0 d\epsilon \bigg[-\sqrt{2\alpha
\epsilon} \,(a_\epsilon +a^\dagger_\epsilon)S_z 
+ \epsilon \, a^\dagger_\epsilon a_\epsilon\bigg],  \label{sbbulla}
\end{eqnarray}

Then, we introduce a logarithmic discretization of the energy interval $[0, 1]$, {\it i.e.}, 
a decreasing set of frequencies $(1, \Lambda^{-1}, \cdots, \Lambda^{-n}, \cdots )$
with $\Lambda>1$ and integer $n$. 
After transforming to a semi-infinite chain, $\widetilde{H}$ reads\cite{Bulla} 
\begin{eqnarray}
H_c&=&{\rm lim}_{N\rightarrow \infty} \Lambda^{-N} H_N, \label{chainmodel} \\
H_N&=& \Lambda^{N} \bigg[ H_S + \sqrt{\alpha} \, \omega_c \, (b_0+b^\dagger_0) 
S_z  \nonumber \\
&&  +\sum^N_{n=0} \epsilon_n b^\dagger_n b_n + 
\sum^{N-1}_{n=0} t_n ( b^\dagger_n b_{n+1}+ h.c.) \bigg].   \label{Hc}
\end{eqnarray}
In Eq.~(\ref{chainmodel}) the spin gets only coupled to the first ($0$th) site of 
the bosonic chain, and the remaining part of the chain is characterized by on-site 
energies $\epsilon_n$ and hopping parameters $t_n$. They
satisfy a set of recursion relations which can be solved numerically. We refer the 
reader to Ref.~\onlinecite{Bulla} for the details of these recursion relations
and their derivations as well as for the precise definition of the $b_n$ boson operators.
It is important to note that both $\epsilon_n$ and $t_n$ decay 
exponentially as $\Lambda^{-n}$ as a result of the logarithmic discretization.
This allows us to solve the model in Eq.~(\ref{chainmodel}) in an iterative 
way: first we diagonalize $H_N$ exactly, keep the $N_s$ lowest energy
levels and then use the recursion relation between $H_{N+1}$ and $H_N$
derived from Eq.~(\ref{Hc}) to diagonalize $H_{N+1}$.  Typically, the result 
converges at large $N \simeq 25$. 
In contrast to the fermionic case, each boson site allows for an infinite number 
of bosons and hence a truncation of the $(N_b+1)$ boson 
states is necessary \cite{DMRG}. Throughout this paper, we use $N_s=100$ and $N_b=8$ (except 
the 0th site for which $N_{b0}=500)$; those values are approximately those used in Ref. 
\onlinecite{Bulla}. We have also used another set of parameters ($N_s=150$ and $N_b=12$) 
and found a good agreement. The
\begin{figure}[h]
\begin{picture}(180,140)
\leavevmode\centering\includegraphics{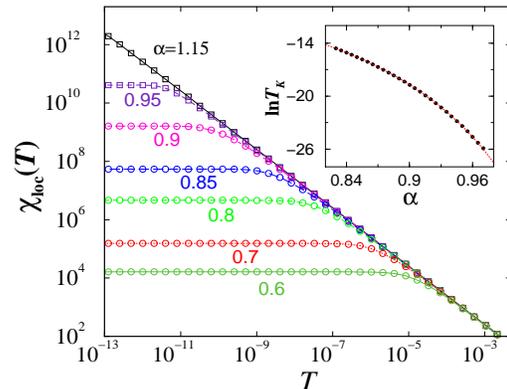}
\end{picture}
\caption{(color online) The local susceptibility $\chi_{\rm loc}(T)$ as a function 
of $T$. The open circles are the numerical data for $T=T_N=1.1\Lambda^{-N}$ and lines 
are guides to the eyes. 
Inset: Logarithm of the Kondo energy scale $T_K$ versus $\alpha$ for $\alpha$ close 
to $\alpha_c$. The solid circles are the numerical data and the dashed line represents 
the fitting curve ln$T_K=-4.67/(1.15-\alpha)$. The NRG parameters, $\Delta$ and 
$\omega_c$, are the same as in Fig.~\ref{Sz}.}
\label{chi} 
\end{figure}
quantities $\langle S_z \rangle(h)$ and $\chi_{\rm loc}(T)=
d \langle S_z \rangle(T)/dh|_{h\rightarrow 0}$ 
 are also calculated iteratively.

{\em NRG Results.}---In Fig.~\ref{Sz} (left) we show the flow of the ground state 
expectation value $\langle S_z \rangle$ as a function of the iteration variable $N$. 
At small $\Delta/\omega_c$, for $\alpha=1.4$ 
and for any $h<0$, after a few iterations we find that $\langle S_z \rangle$ flows 
rapidly to $1/2$, which is in agreement with the localized (dissipative) phase of 
the CL model\cite{CL} as well as with the perturbative RG analysis of the BFKM 
\cite{karyn}. This corresponds to an unscreened local 
spin or a particle which is localized in one of the two allowed levels. 
A strong dissipative Ising coupling with the bosonic bath definitely hinders the Kondo 
effect to develop. In contrast, for $\alpha=0.6$, we show that $\langle S_z \rangle$
flows to saturating values which deviate from $1/2$ and fall drastically
with the decreasing $h$. Moreover, $\langle S_z \rangle$ yields a linear dependence versus 
$h$ around $h=0$ which indicates a fully screened moment: this is a clear signature of the 
Kondo effect. We have fitted the NRG curve for $\alpha=0.6$ (which is sufficiently far from 
$\alpha_c$) with Bethe-Ansatz results applying to the Kondo realm\cite{BA,Buttiker}. 
The dependence of $\langle S_z \rangle$ on $h$ at various $\alpha$ for $\Delta=0.01$ and 
$\omega_c=1$ is displayed in Fig.~\ref{Sz}.
 
We have carefully followed the progressive destruction 
of the delocalized Kondo regime when approaching the quantum phase transition 
at different applied magnetic fields. More specifically, we can infer the Kondo 
temperature $T_K$ from $\langle S_z \rangle$ versus $h$ based on the following 
scaling argument. 
The (Kondo) Fermi liquid type state emerging for $\alpha<\alpha_c$ is embodied by a 
constant (local) spin susceptibility $\chi_{\rm loc}=\partial \langle S_z 
\rangle /\partial h$ which must be identified as\cite{karyn,BA} $\chi_{\rm loc} 
\approx 1/T_K$. For $\alpha$ not too far from $\alpha_c$, we can then extract $T_K$ as
the width of $h$ for $\langle S_z \rangle$ changing from -0.45 to 0.45. This NRG 
procedure is complementary to that from the energy spectrum flow 
\cite{Bulla}. We check that the resulting $T_K$ obeys $\ln T_K \propto 1/(\alpha_c-\alpha)$ 
close to $\alpha_c$ as clearly shown in the inset of Fig.~\ref{chi} for 
$\Delta=0.01$ and $\omega_c=1$ \cite{note3}.
The fact that $T_K$ decreases exponentially fast close to $\alpha_c$ is 
definitely reminiscent of the KT transition. The scaling in the 
inset of Fig.~2 also reveals that for the used NRG parameter $\Lambda=2$, 
$\alpha_c=1.15$. In fact, the NRG result is only exact 
in the limit of $\Lambda\rightarrow 1$. 
Hence, the precise $\alpha_c$ can be obtained by getting $\alpha_c$ for a few 
$\Lambda$ parameters and by extrapolating the result
to $\Lambda \rightarrow 1$.  For $\Delta/\omega_c=0.01$, we reach 
$\alpha_c(\Lambda\rightarrow 1) = 0.98\  (\simeq 1)$. The effect of the 
dissipative coupling becomes weaker when increasing $\Delta$. For each $\Delta$, 
we have repeated the above-mentioned extrapolation procedure and precisely
extracted $\alpha_c$ which basically corresponds to the critical bosonic coupling 
at which the Kondo scale strictly vanishes. 
The resulting $\alpha_c$ as a function of $\Delta$ obeys
$\alpha_c(\Delta)=0.974+0.537\Delta/\omega_c$ which clearly agrees with our expectation from 
the CL mapping. The formation of a  Fermi liquid ground state in the Kondo realm is also clearly 
shown in Fig.~\ref{chi} through the $\chi_{\rm loc}(T)$ versus $T$ plot: For $\alpha<\alpha_c=1.15$, 
$\chi_{\rm loc}(T)$ saturates to a constant for $T<T_K$ whereas for $T$ above $T_K$ 
the Curie's law $\chi_{\rm loc}\propto 1/T$ is well satisfied. At the transition point, we 
clearly observe that
 the $\chi_{\rm loc}(T)\propto 1/T$ behavior persists down to the lowest 
temperatures.

Finally, we would like to briefly comment on the effect of the Ising coupling term 
$J_z$ between the spin and the fermions, originally present
in the heavy fermion systems but absent in the mesoscopic realizations 
under consideration, close to the phase transition point. For $\alpha\rightarrow 1$,
$\varphi \simeq \varphi_s$ and thus the second term in the right hand side of 
Eq.~(\ref{CL}) becomes ${\sqrt{2}}(\frac{1}{4\pi}J_z-\frac{\sqrt{\alpha}}{2}v_s) 
\partial_x\varphi_s(0) S_z$. This immediately leads to a critical $\alpha_c \simeq 1
+ {\cal O}(\Delta) + {\cal O}(J_z)$, which is qualitatively consistent with 
Ref.~\onlinecite{Borda}. 

{\em Conclusion.}---We have demonstrated that the BFKM, with an ohmic bosonic bath and an
Ising-type coupling between the latter and the impurity spin, and the CL model 
belong to the same class of universality. This unambiguously proves the existence of a KT phase 
transition at zero temperature in the former model when 
tuning the Ising dissipative coupling. 
We have chosen to resort to the bosonic NRG to investigate in detail physical quantities like 
the spin magnetization $\langle S_z \rangle$ versus the magnetic field $h$ close to the quantum 
phase transition where Bethe-Ansatz calculations break down. The latter clearly reflects two phases: 
the ``localized'' phase typical of an unscreened moment and the ``delocalized'' Kondo realm 
characterized by a Kondo energy scale which goes to zero exponentially fast close to the KT 
transition. Comparing with the recent fermionic NRG of the BFKM\cite{Borda}, the 
bosonic NRG approach can be extended to the case of a sub-ohmic 
bath\cite{Bulla,MPKW} which might be relevant for the quest of quantum criticality in heavy-fermion 
systems\cite{IngersentN}.

M.-R. Li thanks R.~Bulla for discussions on the bosonic NRG. 
K.L.H. was supported by CIAR, FQRNT, and NSERC. K.L.H. thanks hospitality 
of SPHT Saclay and of KITP (Santa-Barbara) through the ``Quantum phase transition'' workshop, April 
2005 (NSF PHY99-07949).


\begin{thebibliography}{10}
\bibitem{note2} Following the perturbative RG of Ref.~\onlinecite{karyn}, we get the 
critical resistance $R_c=2J_{\perp}N R_K/\nu=R_K/(2K_b)$ where $N=1/(2\pi v_f)$ is the 
fermionic density 
of states and due to the definition of the boson operator $\partial_x\Phi(0)$ 
here $\nu=2$. The perturbative RG also predicts $T_K=\omega_c \exp[-4/(\alpha_c
-\alpha)]$ which resembles the fit of Fig. 2 (inset).
\bibitem{note} From bosonization, $\varphi_s(x) = 
-\sqrt{2\pi\over L}\sum_{q>0}{1\over i\sqrt{q}}(a_qe^{iqx}-h.c.) 
e^{-aq/2}$, where $L$ is the length of the system and $a$ is a typical short-distance
cutoff; note that $[\partial_x\varphi_s(x),\varphi_s(y)]=i2\pi \delta(x-y)$. This 
leads to $\widetilde{H}=H_S+2\sqrt{\alpha}v_s \sqrt{\pi\over L}\sum_{q>0} \sqrt{q}
(a_q+a_q^\dagger) e^{-aq/2} S_z + v_s \sum_{q>0} q a_q^\dagger a_q$. Rescaling 
$q=\epsilon/a$ and equating the high-energy cutoff $\omega_c=v_s/a$, we reach 
Eq.~(\ref{sbbulla}) with $[a_{\epsilon},a_{\epsilon'}^\dagger]=\delta(\epsilon
-\epsilon')$.
\bibitem{note3} The mapping requires $\alpha\rightarrow 1$. 
So to check the result in Eq.~(4), we must examine $T_K$ for $\alpha$ close to $\alpha_c$. 
For $\alpha$ far away from $\alpha_c$ there is a similar scaling $\ln T_K \propto 
1/(\alpha_c-\alpha)$ \cite{Bulla,costi}, but the prefactor
$\ln (\Delta/\omega_c)$ is not equal to -4 \cite{note2}.
\end{thebibliography}
\end{document}